\documentclass[twocolumn,prb,showpacs]{revtex4-1}
\usepackage{epsfig,amsmath,amssymb,color}
\begin{document}

\title{Interplay of charge, spin and lattice degrees of freedom\\
on the spectral properties of the one-dimensional Hubbard-Holstein
model}

\author{A. Nocera$^{1}$, M. Soltanieh-ha$^{1}$, C.A. Perroni$^{2}$, V. Cataudella$^{2}$, and A. E. Feiguin$^{1}$}

\affiliation{$^{1}$Department of Physics, Northeastern University,
Boston MA 02115,
USA\\
$^{2}$ CNR-SPIN and Dipartimento di Fisica, Univ. �Federico II�,
Via Cinthia, Napoli, I-80126, Italy}

\begin{abstract}
We calculate the spectral function of the one dimensional
Hubbard-Holstein model using the time dependent Density Matrix
Renormalization Group (tDMRG), focusing on the regime of large
local Coulomb repulsion, and away from electronic half-filling. We
argue that, from weak to intermediate electron-phonon coupling,
phonons interact only with the electronic charge, and not with the
spin degrees of freedom. For strong electron-phonon interaction,
spinon and holon bands are not discernible anymore and the system
is well described by a spinless polaronic liquid. In this regime,
we observe multiple peaks in the spectrum with an energy
separation corresponding to the energy  of the lattice vibrations
(i.e., phonons). We support the numerical results by introducing a
well controlled analytical approach based on Ogata-Shiba's
factorized wave-function, showing that the spectrum can be
understood as a convolution of three contributions, originating
from charge, spin, and lattice sectors. We recognize and interpret
these signatures in the spectral properties and discuss the
experimental implications.
\end{abstract}

\maketitle
\newpage

\section{Introduction}

In the past two decades, we have witnessed a tremendous
improvement in the energy and momentum resolution of
angle-resolved photoemission spectroscopy (ARPES), which is one of
the most powerful experimental tools for investigating strongly
correlated materials. In particular, recent ARPES spectra of
high-T$_{C}$ cuprates\cite{Lanzara1,Gweon}, alkali-doped
fullerides\cite{Gunnarsson}, and manganites\cite{Lanzara2}, have
shown that the interplay of electron-electron (e-e) and
electron-phonon (e-ph) interactions have an important role in the
qualitative and quantitative understanding of the experiments.

When considering systems of low dimensionality, the situation is
even more complicated. In one dimension (1D), the low-energy
states separate into spin (spinon) and charge (holon) excitations
that move with different velocities and are at different energy
scales \cite{Gogolin,Giamarchibook,Deshpande}. Spin-charge
separation (SCS) has been observed experimentally in semiconductor
quantum wires\cite{auslaender}, organic conductors\cite{Lorenz},
carbon nanotubes\cite{Bockrath}, and atomic chains on
semiconductor surfaces\cite{Blumenstein}. It has also been
predicted that SCS can be achieved in optical lattices of
ultracold atoms\cite{Kollath,Kollath2006,Feiguin2009}. The
phenomenon has been observed in photoemission experiments on
quasi-1D cuprate $SrCuO_{2}$\cite{Kim} and on organic conductor
TTF-TCNQ\cite{Sing}. The coupling to the lattice is considered to
be responsible for the unusual spectral broadening of the spin and
charge peaks observed by ARPES in these quasi-1D materials.
Recently, the interplay between spin, charge, and lattice degrees
of freedom has also been investigated in the family of quasi-1D
cuprates Ca$_{2+5x}$Y$_{2-5x}$Cu$_{5}$O$_{10}$, using high
resolution resonant inelastic x-ray scattering
(RIXS)\cite{DeverauxPRL2013,DeverauxPRB2014}.

In 1D systems and in the absence of e-ph interaction, the spin
excitations are described by a band whose curvature is
proportional to the exchange energy scale $J$, while the charge
excitation dispersion width is comparable to the electronic
hopping ($\simeq 4t$). Moreover, the collective excitation
spectrum of 1D systems presents spectral weight (shadow bands) at
momenta larger than the Fermi momentum $k_{F}$ due to their
Luttinger liquid nature. It is therefore of paramount importance
to study the behavior of the photoemission spectrum of materials
in which it is believed that a strong interaction with the lattice
degrees of freedom is present. In particular, this aspect is not
entirely understood and one expects that the interplay between e-e
and e-ph interactions has a profound effect on SCS and on the
interpretation of the experiments.

The basic lattice model used to describe e-e and e-ph interactions
in 1D is given by the Hubbard-Holstein (HH) Hamiltonian, which
incorporates nearest-neighbor hopping, an on-site Coulomb
repulsion and a linear coupling between the charge density and the
lattice deformation of a dispersionless phonon mode. Within this
model, the electronic spectral properties have been studied by
Ref.\onlinecite{matsueda} and \onlinecite{ning}, where the
adiabatic limit (phonon frequency smaller than the electronic
hopping) is mostly analyzed at half electronic filling in the
regime of weak to intermediate e-ph coupling. In the first paper,
the authors use dynamical density matrix renormalization group
(D-DMRG) and assess the robustness of SCS against e-ph coupling,
interpreting the spectral function as a superposition spectra of
spinless fermions dressed by phonons. In particular, a
peak-dip-hump structure is found, where the dip energy scale is
given by the phonon frequency and originated from the
charge-mediated coupling of phonons and spinons. In the second
paper, the authors use cluster perturbation theory (CPT) and an
optimized phonon approach observing that e-ph coupling mainly
gives rise to a broadening of the holon band, due to the presence
of many adiabatic phonons.

In contrast to these previous studies, in this paper we consider
the case of a finite hole doping (or equivalently electronic
density different from half-filling), a regime that could be
currently accessible in the experiments\cite{DeverauxPRB2014}.
Moreover, we systematically study the spectral properties as a
function of the e-ph coupling and of the phonon frequency,
focusing on the regime where phonon frequency is equal to the
electronic half bandwidth $2t$ (larger than exchange energy $J$).
In order to address this problem, we numerically calculate the
spectral function (photoemission spectrum, PES) of the HH model in
1D using the time-dependent DMRG\cite{tDMRGFeiguin,tDMRGVidal}
(tDMRG). The tDMRG is a robust and unbiased numerical technique
for studying the dynamics of 1D quantum lattice models, that we
apply in the presence of phononic degrees of freedom. We consider
the regime in which a very large Coulomb repulsion is present in
order to avoid competition with other phases such as the CDW
Peierls state\cite{Hardikar2007} and to analyze the effects of a
small exchange energy $J$. One of the main observations is that
the e-ph interaction induces a reduction of the spinon and the
holon band amplitudes, from weak up to intermediate e-ph coupling.
In this case phonons are mainly coupled to the charge degrees of
freedom while the spinon is pretty much unaffected within good
approximation. Eventually, in the strong e-ph coupling regime, one
observes that the separation of spin and charge spectral peaks is
not appreciable anymore being spinon and holon bands merged in one
main band. Moreover, there is a transfer of spectral weight in
side-bands separated from the main band by an energy difference
approximately equal to the phonon frequency. Therefore, for strong
e-ph coupling, the system can be described as a polaronic liquid,
with the spectral weight extended well beyond Fermi momentum
$k_F$.

In order to interpret and understand these results, we develop a
controlled analytical approach to obtain the spectral function. By
construction, this approach is rigorously valid in the presence of
an infinitely large Coulomb repulsion, and a phonon frequency
larger than the electronic hopping. In this regime, we approximate
the ground state as a product of the Ogata-Shiba's
wave-function\cite{Ogata1990} resulting from the exact Bethe
ansatz solution of the $U\rightarrow\infty$ Hubbard model, and a
noninteracting displaced phonon wave-function. This calculation
provides a good qualitative and quantitative agreement with the
tDMRG results in the weak and strong e-ph coupling regime. In this
latter case, the spectral side-bands at intervals of energy
proportional to the phonon frequency are almost coincident within
tDMRG and analytical approaches. Finally, the PES is investigated
with tDMRG decreasing the phonon frequency and exploring also the
adiabatic limit. In this case, we reproduce the results of
Ref.\onlinecite{matsueda} finding the characteristic spectral
peak-dip-hump structure.

The paper is organized as follows: In Sec.II the HH model is
briefly introduced; In Sec.III the method employed to calculate
the spectral function is presented. In Sec.IV, the numerical
results obtained from the tDMRG are discussed and analyzed. In
Sec.V, an analytical method for calculating the spectral function,
its validity, and comparison with the tDMRG results are discussed.
We finally conclude discussing the implications of our results for
the experiments.

\section {The 1D Hubbard-Holstein model}
The Hubbard-Holstein model describes Einstein phonons locally coupled to electrons described
by the Hubbard Hamiltonian. It can be written in
the general form
\begin{eqnarray}
 \label{HH}
H&=&-t \sum\limits_{<i,j>, \sigma}(c^\dagger_{i,\sigma}
c_{j,\sigma} + h.c.) + U \sum\limits_{i} n_{i,\sigma}
n_{i,\bar{\sigma}} \nonumber\\
&+& \omega_0 \sum\limits_i a^\dagger_i a_i + g \omega_0
\sum\limits_{i, \sigma} n_{i\sigma} (a_i+a^\dagger_i),
\end{eqnarray}
where $t$ is the hopping amplitude between nearest neighbor sites
(indicated by $<i,j>$), $U$ is the on-site Coulomb repulsion,
$\omega_0$ is the phonon frequency, $g$ is the e-ph coupling
constant, $c^\dagger_{i,\sigma}$ ($c_{i,\sigma}$) is the standard
electron creation (annihilation) operator on site $i$ with spin
${\sigma}$ ($\bar{\sigma}$ indicates the opposite of $\sigma$),
$n_{i,\sigma}=c^\dagger_{i,\sigma} c_{i,\sigma}$ is the electronic
occupation operator, and $a^\dagger_i$ ($a_i$) is the phonon
creation (annihilation) operator. The Planck constant is set to
$\hbar=1$, the lattice parameter $a=1$, and all of the energies
are in the units of the hopping $t$.

It is well known that the HH model is extremely complicated and
impossible to solve analytically. Its phase diagram and
ground-state static
properties\cite{Hardikar2007,Littlewood2011,Littlewood2012,Perroni2005,Assaad2013,Senechal2011,Deveraux2012,Bauer2010,Bauer2010b,vandenBrink2008,Fabrizio2008,Fehske2008}
have been thoroughly studied in the literature, using different
numerical techniques, including the
DMRG\cite{Tezuka2007,Fehske2010,Fehske2009}. The main difficulty
consists of handling the phononic degrees of freedom, that need to
be described in principle by an infinite dimensional Hilbert space
at every lattice site. Different truncation schemes for the
phononic Hilbert space have been proposed, including the
possibility of using optimal phonon
bases\cite{WhiteJeckelmann98,WhiteJeckelmann99,Perroni2004}.
Still, solving the problem numerically remains remarkably time
consuming, especially for the calculation of the dynamical
properties such as the spectral function. In the next section, the
PES of the 1D HH model is calculated using the tDMRG. The
numerical results are then presented and compared with an
analytical method we introduce in section V.

\section{SPECTRAL FUNCTION WITH \MakeLowercase{t}DMRG}

In order to obtain dynamical properties of 1D quantum lattice
models in the presence of phonons, several techniques such as
dynamical DMRG\cite{matsueda} and exact diagonalization combined
with cluster perturbation theory have been used in the
literature\cite{ning}. In contrast to these approaches, here the
PES is calculated using the tDMRG with Krylov expansion of the
time-evolution
operator\cite{tDMRGManmana,tDMRGSchmitt,cazalilla2002,cazalilla2003,luo}.
The time evolution is computed using $m=400$ DMRG states and the
bare phonon bases are truncated keeping up to 9 phonons per site.
Unless otherwise stated, a lattice with $L=32$ sites, $N=24$
electrons and open boundary conditions is considered. In order to
calculate the PES, we measure the time dependent correlation
function
\begin{equation}\label{PES}
B_{i,j}(t)=i\langle \Psi_{0}| e^{i H t}c^{\dag}_{i}e^{-i H
t}c_{j}|\Psi_{0}\rangle,
\end{equation}
where $|\Psi_{0}\rangle$ is the ground-state of Hamiltonian
(\ref{HH}). $|\Psi_{0}\rangle$ and the ground-state energy are
calculated using static DMRG. Excited states
$|\Psi_{j}\rangle=c_{j}|\Psi_{0}\rangle$ and their time evolution
$|\Psi_{j}(t)\rangle=e^{-i H t}|\Psi_{j}\rangle$ are then
calculated with the tDRMG. Now, since ground-state time evolute is
trivial $\langle \Psi_{0}| e^{i H t}=e^{i E_{gs} t}\langle
\Psi_{0}|$, we thus calculate Eq.(\ref{PES}) simply as
\begin{equation}\label{PES1}
B_{i,j}(t)=i e^{i E_{gs} t}\langle
\Psi_{0}|c^{\dag}_{i}|\Psi_{j}(t)\rangle,
\end{equation}
for $i,j=0,L-1$. Long time evolutions up to $T_{end}=40$ with time
steps of $\Delta t=0.01$ are considered, and $B(k,\omega)$ is
obtained by a space-time Fourier Transform performed using a Hann
window function, giving a broadening of the spectral peaks
approximately given by $\delta\simeq0.25$ (the details of the
procedure are reported in Ref.\onlinecite{tDMRGFeiguin}). Here,
$k$ and $\omega$ are the momentum and energy of the electron.

\begin{centering}
\begin{figure}
\centering \vspace{-.42 in}
{\includegraphics[width=15.2cm,height=12cm,angle=0]{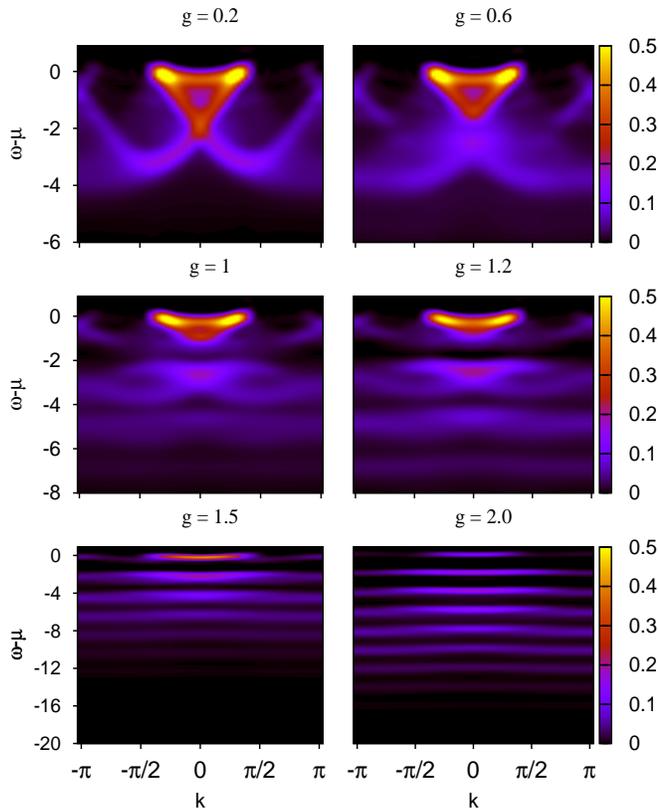}}
\caption{Photoemission spectrum of the HH model calculated with
tDMRG in the antiadiabatic regime ($\omega_{0}=2.0$) for different
e-ph couplings $g=0.2,0.6,1.0,1.2,1.5,2.0$. Here $L=32$ sites,
U=20 and filling N/L=3/4.} \label{fig:PES_contour}
\end{figure}
\end{centering}

\begin{centering}
\begin{figure}
\centering
{\includegraphics[width=8cm,height=7.8cm,angle=0]{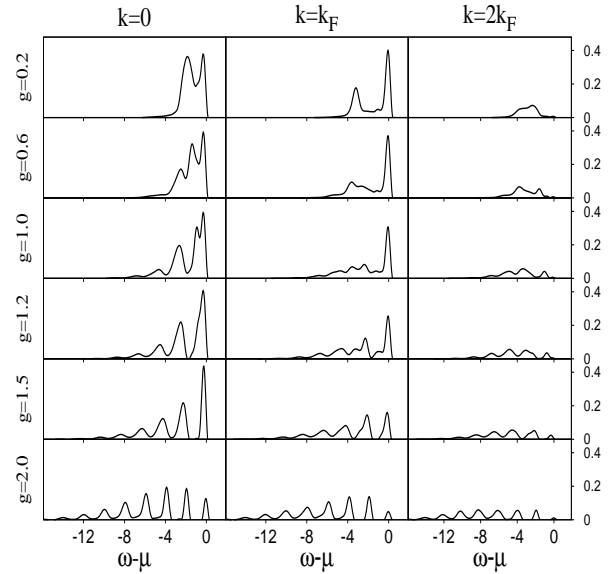}}
\caption{Three cuts at $k=0, k_{F}, 2k_{F}$ of photoemission
spectrum shown in Fig.\ref{fig:PES_contour}.}
\label{fig:Cuts_omega2}
\end{figure}
\end{centering}

\section{\MakeLowercase{t}DMRG RESULTS}
The properties of the PES are analyzed starting from
($\omega_0>1$) and considering in particular $\omega_0=2.0$. In
this regime, Fig.\ref{fig:PES_contour} shows $B(k,\omega)$ from
weak e-ph coupling $g=0.2$ up to strong interaction $g=2.0$.

In order to interpret the results in more detail,
Fig.\ref{fig:Cuts_omega2} is showing three vertical cuts at $k=0$,
$k=k_{F}$ ($k_F=\pi N/2L=0.375\pi$), and $k=2k_{F}$ of the same
spectrum. We note that the spectrum for $g=0.2$ is very similar to
the $g=0$ PES (not shown): it is very clear the presence of SCS,
where the spectral weight concentrated on the spinon and holon
bands forming a triangular spectral structure between $-k_F$ and
$+k_F$ (Fig.\ref{fig:PES_contour}). As expected for a Luttinger
liquid, the shadow bands extend beyond $k_F$. A closer look at PES
in Fig.\ref{fig:Cuts_omega2} in this weak coupling regime, shows
that for $k=k_{F}$ and $k=0$ phonon effects are negligible: one
can observe clearly the higher spinon peak at the top of the
spectrum (at $\omega-\mu\simeq-0.05$ for $k=k_F$), and a shifted
holon peak. The e-ph effects are already present at this weak
coupling for $k=2k_F$, where a shoulder on the left of the main
peak correspondent to the shadow band is visible.

For $g=0.6$, phonon effects come already into play with very
interesting features at all momenta. Looking at
Fig.\ref{fig:PES_contour}, one can observe a reduction of the
spinon and holon bandwidth, as the triangular spectral structure
comprising the spinon and holon bands gets \emph{squeezed}. An
apparent suppression of the spectral weight or gap seems to appear
at $\omega-\mu\simeq-2$ with the formation of a new band ranging
from $\omega-\mu\simeq-2$ to $\omega-\mu\simeq-4$, whose
dispersion resembles those of the holon and shadow bands. The same
characteristics are visible in Fig.\ref{fig:Cuts_omega2} for
$k=0$, where the distance between the spinon peak and the holon
peak is reduced and a side-band at the left of the holon peak is
formed. This new spectral feature seems to originate from the
holon band, while the height of the spinon peak is practically
unchanged going from $g=0.2$ to $g=0.6$.

At $g=1.0$, the progressive reduction of the electronic bandwidth
(both of the spinon and holon bands) is even more evident, and the
triangular spectral structure has almost collapsed. The new band
formed at $g=0.6$ is now separated by a larger gap with respect to
the main spectrum, while the spectral redistribution creates now a
newer side-band whose width is smaller and ranging from
$\omega-\mu\simeq-4$ to $\omega-\mu\simeq-6.2$. As one can see, in
Fig.\ref{fig:Cuts_omega2} for $k=0$, several side-bands separated
in energy by a quantity proportional to $\omega_{0}$ are visible.
The side-bands present no internal structure and suggest that, up
to $g=1.0$, they originate from the holon bands without
contribution from the spinons.

For $g=1.2$, the original triangular feature in the PES is
completely collapsed to a flat structure. Also, if one looks at
Fig.\ref{fig:Cuts_omega2} for $k=0$ and $k=k_F$ for the same value
of $g$, the height of the first spectral peak is dramatically
increased with respect to the case of $g=1.0$. This indicates that
one is entered in the strong e-ph coupling regime where the main
band is followed by many side-bands coming from both holon
\emph{and} spinon bands. This description, as one can see in
Fig.\ref{fig:PES_contour}, is even more evident for $g=1.5$, where
the PES is broken in spectral lines whose weight decreases from
the first structure to the followings and extends beyond the Fermi
momentum $k_F$. Besides, the separation between the holon and the
spinon peak is not discernible anymore, suggesting that the system
is going towards a state that can be described in terms of a
spinless polaronic liquid where the spins are completely
uncorrelated. Indeed, for $g=2.0$, the physics of phonon
side-bands is dominating the PES, observing that the several
spectral structures have a smaller width (compared to $g=1.5$
results), bigger height, and that the first spectral structure has
less weight than the second one. This is reproducing
approximatively a transition to a Gaussian distribution of the
spectral weights typical of the polaronic regime.

In order to investigate further this behavior, we have studied the
ground state density distribution function in momentum space
$n_k=(1/L)\sum_i,je^{-i k (i-j)}\langle c^\dag_i c_j\rangle$ and
the spin-spin correlation function in real space, $\langle
S_z(L/2)S_z(L/2+i)\rangle$. As expected for correlated 1D systems,
the density distribution function in momentum space shown in
Fig.\ref{fig:nk} presents a smooth decrease at the Fermi momentum
$k_F$ for all e-ph coupling values. We point out that the e-ph
coupling reduces the decrease at $k_F$ and, globally, it broadens
the density distribution function. Eventually, for $g=2.0$, one
gets a Gaussian profile with $n_{k=0}\simeq0.45$ and
$n_{k=\pi}\simeq0.325$. In Panel(b) of Fig.\ref{fig:nk}, the
spin-spin correlation function from the center of the chain is
shown. Up to $g=1.5$, spin-spin correlations fast decay as a
function of the distance from the center of the chain with
approximately the same behavior. For $g=2.0$, they decay even
faster, showing evidence that, in the polaronic regime spin
degrees of freedom are completely uncorrelated.

In order to get a better interpretation of the aforementioned
results, in the next section an analytical approach for
calculating the PES will be introduced, explaining the
redistribution of the spectral weight in terms of phonon
side-bands.

\begin{centering}
\centering \hspace{-1cm}
\begin{figure}
{\includegraphics[width=9.5cm,height=9cm,angle=0]{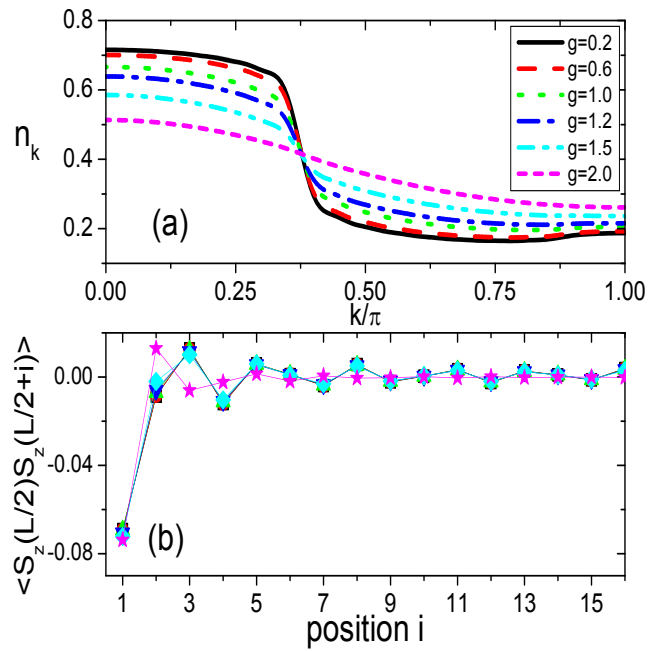}}
\caption{Panel(a) Density distribution function in momentum space
for the same parameter values as in Fig.\ref{fig:PES_contour}.
Solid (black), dashed (red) dotted (green), dashed-dotted (blue),
dashed-dotted-dashed (cyan), short-dashed (magenta), represent
respectively $g=0.2,0.6,1.0,1.2,1.5,2.0$. Panel (b) Spin-spin
correlation function from the center of the chain for the same
parameter values as Panel (a).} \label{fig:nk}
\end{figure}
\end{centering}

\section{ANALYTICAL APPROACH}

In this section we present an analytical method that allows us to
calculate the photoemission part of the spectral function
\begin{equation}
B(k,\omega)=-\frac{1}{\pi} Im G(k,\omega)  \mspace{30mu} for
\mspace{10mu} \omega < \mu, \label{B_green}
\end{equation}
where $G(k,\omega)$ is the electronic retarded single particle
Green's function and $\mu$ is the chemical potential. The method
consists of a variational canonical transformation originally
proposed in Ref.\onlinecite{Zheng} (we refer to it as the ZFA
approach, from the paper of Zheng, Feinberg and Avignon) and then
employed in Ref.\onlinecite{Perroni2003} for calculating the
spectral and optical properties of the spinless Holstein model.
The starting point of the approach is the assumption that, in the
limit of strong e-ph coupling, $U\rightarrow\infty$ and infinite
phonon frequency $\omega_0$, the model is described by spinless
polarons. The ZFA approach, then, extends the polaron formation to
the intermediate e-ph coupling regime, recovering the mean field
solution at zero phonon frequency. The generator of the
variational Lang-Firsov transformation\cite{LangFirsov} is given
by
\begin{equation}
\label{can_trans} T[f,\Delta] = e^{g \sum\limits_j [n_j f +
\Delta](a_j - a^\dagger_j)},
\end{equation}
where $f$ and $\Delta$ are variational parameters. The quantity
$f$ governs the magnitude of the antiadiabatic polaronic effect,
while $\Delta$ represents the lattice distortion proportional to
the average electron density. The transformed Hamiltonian is
\begin{eqnarray}
&&\tilde{H}[f,\Delta]=T^{-1} HT=-t \sum\limits_{<i,j>,
\sigma}(c^\dagger_{i,\sigma} X^\dagger_i X_j c_{j,\sigma} +
h.c.)\nonumber \\
&&+ (U-2 g^2 f^2 \omega_0) \sum\limits_{i} n_{i,\sigma}
n_{i,\bar{\sigma}} + \omega_0 \sum\limits_i a^\dagger_i a_i + L
g^2
\omega_0 \Delta^2\nonumber\\
&&+ g \omega_0 (1-f) \sum\limits_i n_i (a_i+a^\dagger_i) - g
\omega_0 \Delta \sum\limits_i (a_i+a^\dagger_i) \nonumber\\
&& + \tilde\eta \sum\limits_i n_i , \label{trans_HH_1}
\end{eqnarray}
where $L$ is the total number of lattice sites. Here we have
defined a phonon operator $X_i = e^{g f (a_i - a^\dagger_i)}$ and
$\tilde\eta = g^2 \omega_0 f (f-2) + 2 g^2 \omega_0 (f-1) \Delta$.
We leave the technical details of the determination of the
variational parameters $f$ and $\Delta$ in the Appendix
\ref{appendix}. Also, it can be shown easily that the variational
parameter $\Delta$ can be obtained as a function of $f$
($\Delta=(1-f)N/L$), and one is thus left with only one
variational parameter. Once the \emph{optimal} $\tilde f$ is
determined, one can write the transformed Hamiltonian as
\begin{equation}
\label{Htilde_def} \tilde{H}[\tilde{f}] = \tilde{H}_0 + V,
\end{equation}
where $\tilde{H}_0$ is the unperturbed part given by strongly
correlated electrons and non-interacting phonons,
\begin{eqnarray}
&&\tilde{H}_0 [\tilde{f}] = -t e^{-g^2 \tilde{f}^2} \sum\limits_{<i,j>, \sigma}(c^\dagger_{i,\sigma} c_{j,\sigma} + h.c.) \nonumber\\
&&+(U-2 g^2 \tilde{f}^2 \omega_0) \sum\limits_{i} n_{i,\sigma}
n_{i,\bar{\sigma}} + \omega_0 \sum\limits_i a^\dagger_i a_i +
\eta N \nonumber\\
&&- g \omega_0 (1-\tilde{f}) \frac{N}{L} \sum\limits_i
(a_i+a^\dagger_i) + g^2 \omega_0 (1-\tilde{f})^2 \frac {N^2}{L},
\nonumber\\
\end{eqnarray}
with $\eta = g^2 \omega_0 \tilde{f} (\tilde{f}-2) - 2 g^2 \omega_0
(1-\tilde{f})^2 N/L$, while $V$ is a many-body interaction
\begin{eqnarray}
\label{V_def} V &=& -t \sum\limits_{<i,j>,
\sigma}[c^\dagger_{i,\sigma} (X^\dagger_i X_j - e^{-g^2
\tilde{f}^2}) c_{j,\sigma} + h.c.] \\&&+ g \omega_0 (1-\tilde{f})
\sum\limits_i n_i (a_i+a^\dagger_i).\nonumber
\end{eqnarray}
The PES is now calculated approximately by neglecting the
perturbation $V$. One can use perturbation theory and consider the
effect of $V$ in higher orders of perturbation after the
calculation of the PES, but in this paper we are only taking the
zeroth order into account. In fact, the determination of the
optimal parameter $\tilde{f}$ is meant to minimize the error
produced by neglecting the interaction term $V$ from the
Hamiltonian Eq.(\ref{Htilde_def}). The unperturbed Hamiltonian
$\tilde{H}_0 [\tilde{f}]$ still contains information about
interacting terms in the original Hamiltonian Eq.(\ref{HH}), since
all the parameters of $\tilde{H}_0 [\tilde{f}]$ are renormalized
by our variational technique. Indeed, $\tilde{H}_0 [\tilde{f}]$
consists of free phonons and a Hubbard model with a hopping
$\tilde{t}$ and an on-site repulsion $\tilde{U}$ renormalized by
the e-ph interaction
\begin{equation}
\label{t_best}
 \tilde{t}=t e^{-g^2 \tilde{f}^2}, \tilde{U}=U-2 g^2 \tilde{f}^2 \omega_0.
\end{equation}
The PES is now evaluated in the Lehmann representation
\begin{eqnarray}
\label{B_ck}
B(k,\omega)&=&\sum_{\{n_{ph}\},z,\sigma}|\langle|\{n_{ph}\},z,N-1|c_{k,\sigma}|\{0_{ph}\},gs,N \rangle |^2 \nonumber \\
&\times& L\delta (\omega - E_{gs}^N + E_z^{N-1} ),
\end{eqnarray}
where $c_{k,\sigma}$ destroys an electron with momentum $k$ and
spin $\sigma$ ($c_{j,{\sigma}}=\frac{1}{\sqrt{L}} \sum_{k'} e^{i
k' j} c_{k',\sigma}$), $N$ is the total number of electrons, and
$z$ the final state with $N-1$ electrons. $E_{z}^{N-1}$ represents
the total energy of the final state, $|\{n_{ph}\},z,N-1\rangle$,
where a generic phonon contribution is included, and $E_{gs}^N $
describes the energy of the ground state of the original
Hamiltonian (\ref{HH}) with $N$ electrons. Since Einstein phonons
carry no momentum, we can impose the momentum conservation with
the term $\delta_{k,P_{gs}^N-P_{z}^{N-1}}$ to reduce
Eq.(\ref{B_ck}) to a calculation involving only site $0$ in the
real space and one phonon mode at that site
\begin{eqnarray}
\label{B_c0}
B(k,\omega)&=&\sum_{\{n_{ph}\},z,\sigma}|\langle |\{n_{ph}\},z,N-1 |c_{0,\sigma}|\{0_{ph}\},gs,N \rangle |^2 \nonumber\\
&\times& L\delta (\omega - E_{gs}^N + E_{z}
^{N-1})\delta_{k,P_{gs}^N-P_z^{N-1}} .
\end{eqnarray}
Up to here, no assumptions have been made on the spectral function
and this general form is extremely complex.
However, in the basis of $\tilde{H_0}$, the wave-function is
trivially separated into phonon and electronic parts. In the limit
of $\tilde{U} >> \tilde{t}$ one can use Ogata-Shiba's
factorization\cite{Ogata1990} to show that the electronic
wave-function itself is split into spin and charge parts. The
total wave-function can be written as
\begin{equation}
|\mathrm{\psi}\rangle=|\phi\rangle\otimes |\chi\rangle \otimes
|\{n_{ph}\}\rangle. \label{o_sh}
\end{equation}

\begin{centering}
\begin{figure}
\centering  \vspace{-.42 in}
{\includegraphics[width=15.2cm,height=12cm,angle=0]{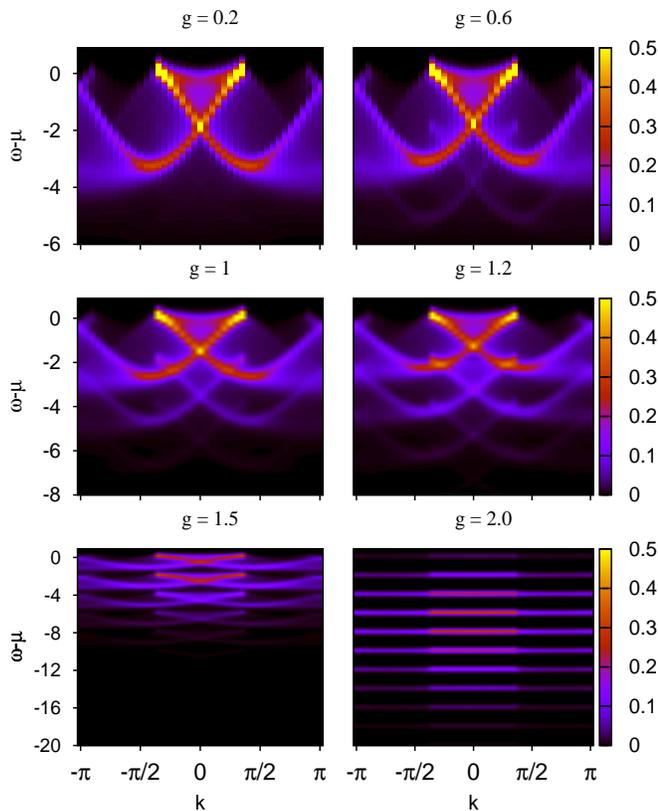}}
\caption{(Color online) Photoemission spectrum calculated with
analytical method for the same frequency and e-ph coupling values
considered in Fig.\ref{fig:PES_contour}. Here $L=40$ sites, $U=20$
and filling $N/L=3/4$.} \label{fig:PES_contour_SEMI}
\end{figure}
\end{centering}

The first part, $|\phi\rangle$, describes spinless charges,
$|\chi\rangle$ is the spin wave-function that corresponds to a
``squeezed'' chain of $N$ spins, where all the unoccupied sites
have been removed, and $|\{n_{ph}\}\rangle$ is given by the
product of $L$ separate non-interacting phononic wave-functions,
each one containing an integer number of phonons
$(|\{n_{ph}\}\rangle=|\{n_{ph}^0\}\rangle \otimes
|\{n_{ph}^1\}\rangle \otimes ... |\{n_{ph}^{L-1}\}\rangle).$ In
this limit, charge, spin, and lattice degrees of freedom are
governed by independent Hamiltonians
\begin{equation}
\label{Htilde0} \tilde{H}_0[\tilde{f}] = \tilde{H}_{charge} +
\tilde{H}_{spin} + \tilde{H}_{phonon}.
\end{equation}
Due to this simplification, we are now able to tackle the problem
and calculate the PES. Indeed, operator $c_{0,\sigma}$ after the
polaron transformation will look like $c_{0,\sigma} X_0$.
Moreover, by using the factorized wave-function and separating
spin and charge operators, $c_{0,\sigma} X_0=Z_{0,\sigma} b_0
X_0$, the spectral function can be expressed as a convolution
\begin{equation}
\label{B_conv} B (k,\omega) = \sum\limits_{\omega ', Q, \sigma}
D_{\sigma} (Q,\omega ') B_Q (k,\omega - \omega ')
\end{equation}
where $D_{\sigma} (Q,\omega)$ is the spin spectral function with
momentum $Q$, and
\begin{eqnarray}
\label{BQ_X} &&B_Q (k,\omega)= L\sum\limits_{ \{ I \} } \{
|\langle\psi_{L,Q}^{N-1}\{I\} | b_0 | \psi_{L,\pi}^{N,gs}\rangle|^2\\
&\times& \sum\limits_{\tilde{n}} |\langle\tilde{n} | X_0 |
0\rangle|^2 \delta (\omega-E^N_{gs}+E^{N-1}_z+\tilde{n}
\omega_0)\nonumber\\&& \delta_{k,P^N-P^{N-1}}\},\nonumber
\end{eqnarray}
describes the charge and phonon parts. By following the approach
introduced in Ref.\onlinecite{Penc1997b}, one can calculate numerically both
$D_{\sigma} (Q,\omega)$ and $B_Q (k,\omega)$.
\begin{centering}
\begin{figure}
\centering
{\includegraphics[width=8cm,height=7.8cm,angle=0]{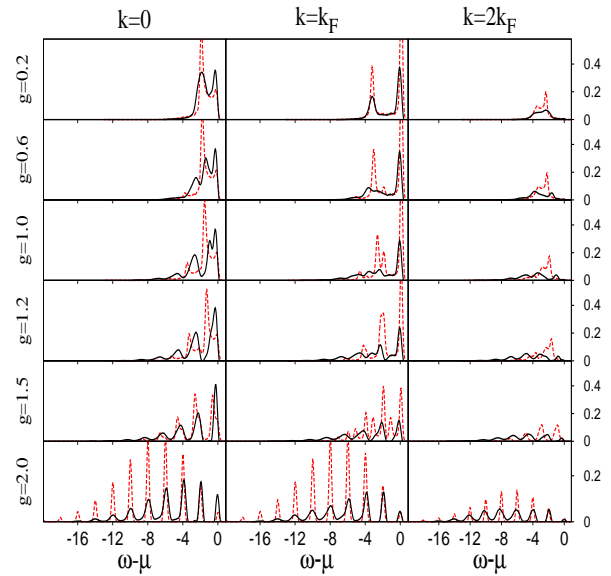}}
\caption{Three cuts at $k=0, k_{F}, 2k_{F}$ of photoemission
spectrum calculated with the tDMRG (solid (black) line, shown also
in Fig.\ref{fig:Cuts_omega2}) and using the ZFA approach (dashed
(red) line).} \label{fig:Cuts_omega2_SEMI}
\end{figure}
\end{centering}

Fig.\ref{fig:PES_contour_SEMI} shows the PES calculated with the
ZFA approach in the antiadiabatic regime, for the same regime of
parameters of Fig.\ref{fig:PES_contour}. In analogy with the tDMRG
results, we also show three vertical cuts of the spectrum at
$k=0$, $k=k_F$, $k=2k_F$ in Fig.\ref{fig:Cuts_omega2_SEMI} (dashed
(red) line). It is important to point out that, even within the
ZFA approach, a broadening of the spectral peaks of the order of
$\delta\simeq0.25$ has been used.

As stated at the beginning of this section, one expects that the
ZFA approach is a good approximation of the results in the regime
where $\tilde{U} >> \tilde{t}$ and $\omega_0 > \tilde{t}$. Also,
the optimized polaronic parameter $\tilde f$ (see Appendix
\ref{appendix}) is describing the degree of polaron formation,
that is the amount of spectral weight redistribution in phonon
side-bands. In general, for $\tilde f=1$ one has well defined
polarons, while, for $\tilde f=0$, the unitary transformation,
Eq.\ref{can_trans}, becomes trivially the identity. As one can
observe in Fig.\ref{fig:f_best_energy} (Appendix \ref{appendix}),
for the set of parameters used in this paper, $U=20$ and
$\omega_0=2.0$, $\tilde f$ assumes a value of $0.4$ for $g=0.2$
increasing slightly up to $0.5$ for $g=1.2$, pointing out that
strong Coulomb repulsion and the large phonon frequency already
give a sizeable effect from weak to intermediate e-ph couplings.
In particular, as one can see in the top row of panels of
Fig.\ref{fig:Cuts_omega2}, for $g=0.2$ a very good agreement
between ZFA and the tDMRG results is obtained. This characteristic
is also evident at all momenta if one looks at the upper left
panel of Fig.\ref{fig:PES_contour} and
Fig.\ref{fig:PES_contour_SEMI}.
\begin{centering}
\begin{figure}
\centering \hspace{-.45 in}
{\includegraphics[width=9.6cm,height=7.5cm,angle=0]{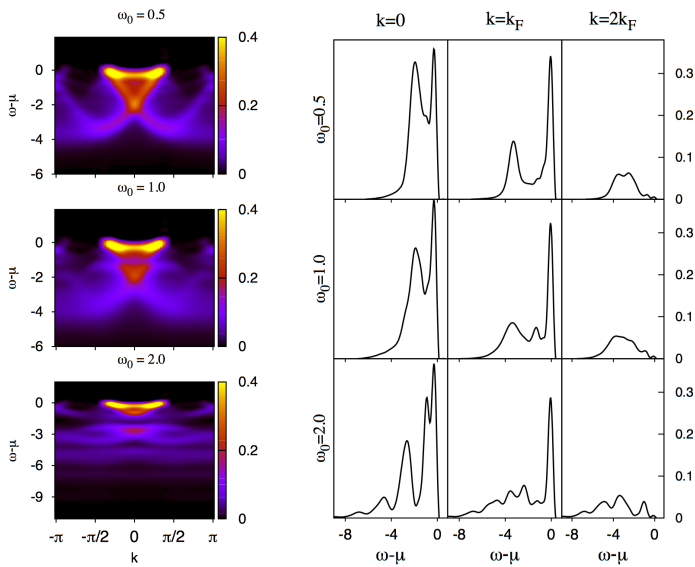}}
\caption{Panel(a) Photoemission spectrum at e-ph couplings
$g=1.0$, for different phonon frequencies
$\omega_{0}=0.5,1.0,2.0$. Panel(b) Cut at $k=0$, $k=k_{F}$, and
$k=2k_{F}$ of photoemission spectrum reported in Panel (a).}
\label{fig:g1}
\end{figure}
\end{centering}

Noticeable differences between the ZFA approach and the tDMRG
results can be observed in the intermediate e-ph coupling regime
($g=0.6,1.0,1.2$). In this case, the ZFA approach is qualitatively
reproducing the reduction of the spinon and holon bandwidths,
which are parametrized by the renormalized hopping parameter
$\tilde t=t e^{-g^2{\tilde f}^2}$ in the Hamiltonian $H_{0}[\tilde
f]$, Eq.(\ref{Htilde_def}). Moreover, while reproducing correctly
the spectral position of the phonon side-bands, the ZFA approach
provides access to their internal structure, showing that the
separation between the holon and spinon peaks is still well
defined.

At strong e-ph coupling, one has $\tilde f=0.675$ for $g=1.5$ and
$\tilde f=0.975$ for $g=2.0$, observing a large polaronic effect.
In these cases, the PES calculated within the ZFA approach
provides the same number of phonon side bands with widths and
heights of the same order of magnitude of the tDMRG results. As in
the tDMRG, the internal structure of the phonon side-bands is
lost, while a clear Gaussian-like distribution of the spectral
weight is observable for $g=2.0$. Strikingly, the ZFA approach is
giving qualitatively the same non-zero spectral weight
distribution at momenta larger than $k_F$, confirming that, in
this case, the system can be described as a polaron liquid. It is
important to observe finally that our analytical approach provides
a shift of the chemical potential given by the quantity ${\tilde
\eta} = g^2 \omega_0 {\tilde f} ({\tilde f}-2) + 2 g^2 \omega_0
({\tilde f}-1) \Delta$. The optimal shift is in total agreement
with tDMRG results in the whole range of e-ph couplings.

We can now briefly discuss the results described above making a
contact with the experiments described in
Ref.\onlinecite{DeverauxPRL2013}. In this paper, the authors
measure the RIXS spectra of a family quasi 1D cuprates
$Ca_{2+5x}Y_{2-5x}Cu_{5}O_{10}$, an insulating system that can be
doped over a wide range of hole concentrations. The experiment
reveals a $70$ meV phonon (energy larger than the typical transfer
hopping $t$ along chains in quasi 1-D cuprates) strongly coupled
to the electronic state at different hole dopings. It is found
that the spectral weight of phonon excitations in the RIXS
spectrum is directly dependent on the e-ph coupling strength and
doping, producing multiple peaks in the spectrum with an energy
separation corresponding to the energy of the quanta of the
lattice vibrations, in a fashion similar to what we obtain in the
present paper. We believe that, even in ARPES spectra of these
materials, phonon side-bands structures in the PES could be
observable.

\section{\MakeLowercase{t}DMRG RESULTS FOR INTERMEDIATE \MakeLowercase{e-ph} COUPLING}

In this section, we extend the analysis by discussing tDMRG
results for intermediate e-ph coupling $g=1.0$, as a function of
the phonon frequency $\omega_{0}$. The results are shown in
Fig.\ref{fig:g1}. For $\omega_{0}=0.5$, we observe a behavior
different from that discussed in the previous section. For
instance, at $k=0$, a dip structure at the left side of the spinon
peak is shifted by a quantity equal to $\omega_{0}=0.5$,
reproducing qualitatively the results discussed in
Ref.\onlinecite{matsueda}. In Ref.\onlinecite{matsueda}, the
interpretation of results starts from the consideration that, in
absence of e-ph coupling, according to the Bethe ansatz solution
the PES is constructed by a superposition of a set of holon
dispersions forming a cosine band with width $4t$. Moreover, each
holon dispersion is characterized by one spinon momentum. In the
presence of e-ph interaction, due to spin-charge separation each
holon couples with phonons independently and the PES is
interpreted as a spectrum of spinless electron dressed by Einstein
phonons. This generates a split of the holon dispersion that is
away from the top of the spectrum by a energy interval equal to
$\omega_0$, and a transfer of spectral weight to high energy
giving a characteristic peak-dip-hump structure. Our results are
consistent with this picture, confirming that spin-charge
separation is robust in this regime. Actually, in contrast to what
discussed in the previous section for $\omega_0=2.0$, where
polaronic effects dominate, when the phonon frequency is smaller
than the hopping $t$, the e-ph coupling effect gives rise to a dip
in between the holon and spinon peak. Moreover, this spectral dip
structure is furthermore shifted if $\omega_{0}$ is increased to
$\omega_{0}=1.0$ (See panel(b) of Fig.\ref{fig:g1} for
$\omega_0=1.0$ and $k=0$). In this case, our data shows also a
shoulder on the left side of the holon peak, in contrast to what
found in Ref.\onlinecite{matsueda}. In our calculation, this
feature can be interpreted as the onset of phonon side-bands.
Increasing the phonon frequency to $\omega_{0}=2.0$, several
sidebands in the PES are found as discussed in the previous
section. Interestingly, at $k=k_F$, instead of a dip, we find a
peak separated from the spinon band by a energy distance equal to
$\omega_{0}$. Eventually, at the Fermi momentum and for larger
frequencies, these features become part of the first and the
higher side-bands.

\section{CONCLUSION}
We have studied the spectral function of the 1D HH model using the
tDMRG, in the limit of large Coulomb repulsion, and away from
electronic half-filling. The entire range of e-ph coupling, from
weak to strong coupling $g$, has been analyzed. Our results
indicate that, from weak to intermediate $g$, SCS is robust
against e-ph coupling: the phonons couple mainly with charge
degrees of freedom, leaving the spinon band almost unaffected. For
sufficiently strong e-ph interaction, the PES weight is
redistributed in phonon side-bands, and the spinon and holon
spectral features are not discernable anymore. In this regime, we
support the numerical tDMRG results with an analytical variational
calculation, approximating the wave-function as a convolution of
charge, spin and phonon parts. In this case, a very good
qualitative and quantitative agreement is obtained, and the system
can be described as a polaronic liquid, with non-zero spectral
weight at momenta larger than the Fermi momentum.

\section{ACKNOWLEDGMENTS}
A.E.F. acknowledges NSF support through grant DMR-1339564. A. N.
thanks Lev Vidmar for useful comments.

\appendix

\section{VARIATIONAL CALCULATION OF THE PARAMETER $f$} \label{appendix}

\begin{centering}
\begin{figure}
\centering
{\includegraphics[width=9cm,height=7.8cm,angle=0]{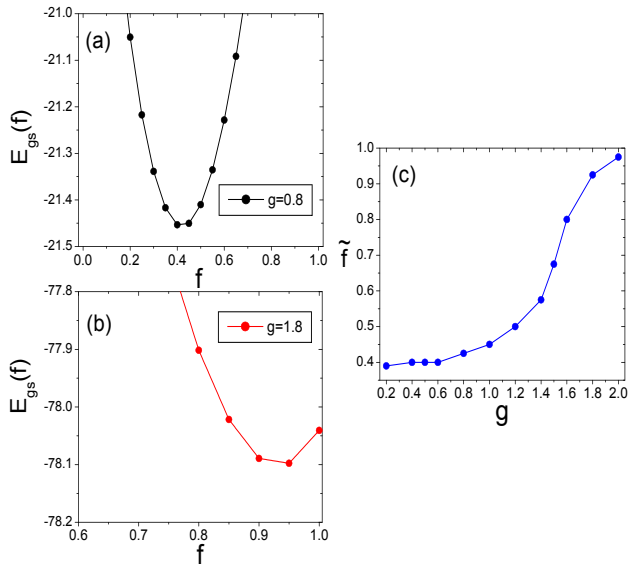}}
\caption{Panels (a) and (b) The ground-state energy as a function
of $f$ in order to get $\tilde{f}$. Panel (c) $\tilde{f}$ as a
function of $g$ for filling N/L=0.75} \label{fig:f_best_energy}
\end{figure}
\end{centering}
In this appendix we determine the variational parameters $f$ and
$\Delta$ appearing in the transformed Hamiltonian
Eq.(\ref{trans_HH_1}) of the main text. An effective electronic
Hamiltonian, $H_{eff}$, is used, which is obtained by averaging
Eq.(\ref{trans_HH_1}) on the phononic vacuum of the transformed
Hilbert space, $H_{eff} [f,\Delta] = \langle
O_{ph}|\tilde{H}|O_{ph}\rangle$
\begin{eqnarray}\label{Heff_def}
&&H_{eff} [f,\Delta]=-t e^{-g^2 f^2} \sum\limits_{<i,j>,
\sigma}(c^\dagger_{i,\sigma} c_{j,\sigma} + h.c.)\\&&+ (U-2 g^2
f^2 \omega_0) \sum\limits_{i} n_{i,\sigma} n_{i,\bar{\sigma}} +
\eta \sum\limits_i n_i  + L g^2 \omega_0 \Delta^2. \nonumber
\end{eqnarray}
The parameter $\Delta$ is simply obtained by using the
Hellmann-Feynman theorem
\[
\frac {\partial}{\partial \Delta} \langle gs|H_{eff}|gs\rangle = 0
\Rightarrow \Delta = (1-f) \frac {N}{L},
\]
where $N$ is the total number of electrons, $N/L$ is the
electronic density, and $|gs \rangle$ is the ground state of
$H_{eff}$. Now we are left only with the determination of the
parameter $f$, which will be found by solving the Hamiltonian
\begin{eqnarray}
\label{Heff_f} &&H_{eff} [f]=-t e^{-g^2 f^2} \sum\limits_{<i,j>,
\sigma}(c^\dagger_{i,\sigma} c_{j,\sigma} + h.c.) \\
&&+ (U-2 g^2 f^2 \omega_0) \sum\limits_{i} n_{i,\sigma}
n_{i,\bar{\sigma}} + g^2 \omega_0 (1-f)^2 N^2/L + \eta N,
\nonumber
\end{eqnarray}
by using the static DMRG and minimizing the ground-state energy of
this new Hamiltonian $H_{eff}$ as a function of $f$. For each set
of values $U$, $t$, $g$, and $\omega_0$, considered in the
original Hamiltonian, we get an optimal polaronic parameter
$\tilde{f}$. In the panels (a) and (b) of
Fig.\ref{fig:f_best_energy}, the ground-state energy of
$H_{eff}[f]$ as a function of $f$ for two different values of e-ph
coupling $g=0.8$ and $g=1.8$, $N/L=0.75$, and $\omega_0=2.0$ is
shown. For $g=1.8$  the value of $\tilde{f}$ obtained is close to
$0.8$ meaning that for these sets of parameters the system is near
the polaronic regime, that ideally is expected to be reached for
stronger e-ph coupling and phonon frequency. In panel (c) of
Fig.\ref{fig:f_best_energy}, the optimal polaronic parameter
$\tilde{f}$ as a function of e-ph is shown as discussed in the
main text.


\begin{thebibliography}{50}
\expandafter\ifx\csname natexlab\endcsname\relax\def\natexlab#1{#1}\fi
\expandafter\ifx\csname bibnamefont\endcsname\relax
  \def\bibnamefont#1{#1}\fi
\expandafter\ifx\csname bibfnamefont\endcsname\relax
  \def\bibfnamefont#1{#1}\fi
\expandafter\ifx\csname citenamefont\endcsname\relax
  \def\citenamefont#1{#1}\fi
\expandafter\ifx\csname url\endcsname\relax
  \def\url#1{\texttt{#1}}\fi
\expandafter\ifx\csname urlprefix\endcsname\relax\def\urlprefix{URL }\fi
\providecommand{\bibinfo}[2]{#2}
\providecommand{\eprint}[2][]{\url{#2}}

\bibitem[{\citenamefont{Lanzara et~al.}(2001)\citenamefont{Lanzara, Bogdanov,
  Zhou, Kellar, Feng, Lu, Yoshida, Eisaki, Fujimori, Kishio et~al.}}]{Lanzara1}
\bibinfo{author}{\bibfnamefont{A.}~\bibnamefont{Lanzara}},
  \bibinfo{author}{\bibfnamefont{P.}~\bibnamefont{Bogdanov}},
  \bibinfo{author}{\bibfnamefont{X.}~\bibnamefont{Zhou}},
  \bibinfo{author}{\bibfnamefont{S.}~\bibnamefont{Kellar}},
  \bibinfo{author}{\bibfnamefont{D.}~\bibnamefont{Feng}},
  \bibinfo{author}{\bibfnamefont{E.}~\bibnamefont{Lu}},
  \bibinfo{author}{\bibfnamefont{T.}~\bibnamefont{Yoshida}},
  \bibinfo{author}{\bibfnamefont{H.}~\bibnamefont{Eisaki}},
  \bibinfo{author}{\bibfnamefont{A.}~\bibnamefont{Fujimori}},
  \bibinfo{author}{\bibfnamefont{K.}~\bibnamefont{Kishio}},
  \bibnamefont{et~al.}, \bibinfo{journal}{Nature}
  \textbf{\bibinfo{volume}{412}}, \bibinfo{pages}{510} (\bibinfo{year}{2001}).

\bibitem[{\citenamefont{Gweon et~al.}(2004)\citenamefont{Gweon, Sasagawa, Zhou,
  Graf, Takagi, Lee, and Lanzara}}]{Gweon}
\bibinfo{author}{\bibfnamefont{G.-H.} \bibnamefont{Gweon}},
  \bibinfo{author}{\bibfnamefont{T.}~\bibnamefont{Sasagawa}},
  \bibinfo{author}{\bibfnamefont{S.}~\bibnamefont{Zhou}},
  \bibinfo{author}{\bibfnamefont{J.}~\bibnamefont{Graf}},
  \bibinfo{author}{\bibfnamefont{H.}~\bibnamefont{Takagi}},
  \bibinfo{author}{\bibfnamefont{D.-H.} \bibnamefont{Lee}}, \bibnamefont{and}
  \bibinfo{author}{\bibfnamefont{A.}~\bibnamefont{Lanzara}},
  \bibinfo{journal}{Nature} \textbf{\bibinfo{volume}{430}},
  \bibinfo{pages}{187} (\bibinfo{year}{2004}).

\bibitem[{\citenamefont{Gunnarsson}(1997)}]{Gunnarsson}
\bibinfo{author}{\bibfnamefont{O.}~\bibnamefont{Gunnarsson}},
  \bibinfo{journal}{Rev. Mod. Phys.} \textbf{\bibinfo{volume}{69}},
  \bibinfo{pages}{575} (\bibinfo{year}{1997}).

\bibitem[{\citenamefont{Lanzara et~al.}(1998)\citenamefont{Lanzara, Saini,
  Brunelli, Natali, Bianconi, Radaelli, and Cheong}}]{Lanzara2}
\bibinfo{author}{\bibfnamefont{A.}~\bibnamefont{Lanzara}},
  \bibinfo{author}{\bibfnamefont{N.}~\bibnamefont{Saini}},
  \bibinfo{author}{\bibfnamefont{M.}~\bibnamefont{Brunelli}},
  \bibinfo{author}{\bibfnamefont{F.}~\bibnamefont{Natali}},
  \bibinfo{author}{\bibfnamefont{A.}~\bibnamefont{Bianconi}},
  \bibinfo{author}{\bibfnamefont{P.}~\bibnamefont{Radaelli}}, \bibnamefont{and}
  \bibinfo{author}{\bibfnamefont{S.-W.} \bibnamefont{Cheong}},
  \bibinfo{journal}{Phys. Rev. Lett.} \textbf{\bibinfo{volume}{81}},
  \bibinfo{pages}{878} (\bibinfo{year}{1998}).

\bibitem[{\citenamefont{Gogolin et~al.}(2004)\citenamefont{Gogolin, Nersesyan,
  and Tsvelik}}]{Gogolin}
\bibinfo{author}{\bibfnamefont{A.~O.} \bibnamefont{Gogolin}},
  \bibinfo{author}{\bibfnamefont{A.~A.} \bibnamefont{Nersesyan}},
  \bibnamefont{and} \bibinfo{author}{\bibfnamefont{A.~M.}
  \bibnamefont{Tsvelik}}, \emph{\bibinfo{title}{Bosonization and strongly
  correlated systems}} (\bibinfo{publisher}{Cambridge University Press},
  \bibinfo{year}{2004}).

\bibitem[{\citenamefont{Giamarchi}(2004)}]{Giamarchibook}
\bibinfo{author}{\bibfnamefont{T.}~\bibnamefont{Giamarchi}},
  \emph{\bibinfo{title}{Quantum Physics in One Dimension}}
  (\bibinfo{publisher}{Clarendon Press, Oxford}, \bibinfo{year}{2004}).

\bibitem[{\citenamefont{Deshpande et~al.}(2010)\citenamefont{Deshpande,
  Bockrath, Glazman, and Yacoby}}]{Deshpande}
\bibinfo{author}{\bibfnamefont{V.~V.} \bibnamefont{Deshpande}},
  \bibinfo{author}{\bibfnamefont{M.}~\bibnamefont{Bockrath}},
  \bibinfo{author}{\bibfnamefont{L.~I.} \bibnamefont{Glazman}},
  \bibnamefont{and} \bibinfo{author}{\bibfnamefont{A.}~\bibnamefont{Yacoby}},
  \bibinfo{journal}{Nature} \textbf{\bibinfo{volume}{464}},
  \bibinfo{pages}{209} (\bibinfo{year}{2010}).

\bibitem[{\citenamefont{Auslaender et~al.}(2005)\citenamefont{Auslaender,
  Steinberg, Yacoby, Tserkovnyak, Halperin, Baldwin, Pfeiffer, and
  West}}]{auslaender}
\bibinfo{author}{\bibfnamefont{O.}~\bibnamefont{Auslaender}},
  \bibinfo{author}{\bibfnamefont{H.}~\bibnamefont{Steinberg}},
  \bibinfo{author}{\bibfnamefont{A.}~\bibnamefont{Yacoby}},
  \bibinfo{author}{\bibfnamefont{Y.}~\bibnamefont{Tserkovnyak}},
  \bibinfo{author}{\bibfnamefont{B.}~\bibnamefont{Halperin}},
  \bibinfo{author}{\bibfnamefont{K.}~\bibnamefont{Baldwin}},
  \bibinfo{author}{\bibfnamefont{L.}~\bibnamefont{Pfeiffer}}, \bibnamefont{and}
  \bibinfo{author}{\bibfnamefont{K.}~\bibnamefont{West}},
  \bibinfo{journal}{Science} \textbf{\bibinfo{volume}{308}},
  \bibinfo{pages}{88} (\bibinfo{year}{2005}).

\bibitem[{\citenamefont{Lorenz et~al.}(2002)\citenamefont{Lorenz, Hofmann,
  Gr{\"u}ninger, Freimuth, Uhrig, Dumm, and Dressel}}]{Lorenz}
\bibinfo{author}{\bibfnamefont{T.}~\bibnamefont{Lorenz}},
  \bibinfo{author}{\bibfnamefont{M.}~\bibnamefont{Hofmann}},
  \bibinfo{author}{\bibfnamefont{M.}~\bibnamefont{Gr{\"u}ninger}},
  \bibinfo{author}{\bibfnamefont{A.}~\bibnamefont{Freimuth}},
  \bibinfo{author}{\bibfnamefont{G.}~\bibnamefont{Uhrig}},
  \bibinfo{author}{\bibfnamefont{M.}~\bibnamefont{Dumm}}, \bibnamefont{and}
  \bibinfo{author}{\bibfnamefont{M.}~\bibnamefont{Dressel}},
  \bibinfo{journal}{Nature} \textbf{\bibinfo{volume}{418}},
  \bibinfo{pages}{614} (\bibinfo{year}{2002}).

\bibitem[{\citenamefont{Bockrath et~al.}(1999)\citenamefont{Bockrath, Cobden,
  Lu, Rinzler, Smalley, Balents, and McEuen}}]{Bockrath}
\bibinfo{author}{\bibfnamefont{M.}~\bibnamefont{Bockrath}},
  \bibinfo{author}{\bibfnamefont{D.~H.} \bibnamefont{Cobden}},
  \bibinfo{author}{\bibfnamefont{J.}~\bibnamefont{Lu}},
  \bibinfo{author}{\bibfnamefont{A.~G.} \bibnamefont{Rinzler}},
  \bibinfo{author}{\bibfnamefont{R.~E.} \bibnamefont{Smalley}},
  \bibinfo{author}{\bibfnamefont{L.}~\bibnamefont{Balents}}, \bibnamefont{and}
  \bibinfo{author}{\bibfnamefont{P.~L.} \bibnamefont{McEuen}},
  \bibinfo{journal}{Nature} \textbf{\bibinfo{volume}{397}},
  \bibinfo{pages}{598} (\bibinfo{year}{1999}).

\bibitem[{\citenamefont{Blumenstein et~al.}(2011)\citenamefont{Blumenstein,
  Sch{\"a}fer, Mietke, Meyer, Dollinger, Lochner, Cui, Patthey, Matzdorf, and
  Claessen}}]{Blumenstein}
\bibinfo{author}{\bibfnamefont{C.}~\bibnamefont{Blumenstein}},
  \bibinfo{author}{\bibfnamefont{J.}~\bibnamefont{Sch{\"a}fer}},
  \bibinfo{author}{\bibfnamefont{S.}~\bibnamefont{Mietke}},
  \bibinfo{author}{\bibfnamefont{S.}~\bibnamefont{Meyer}},
  \bibinfo{author}{\bibfnamefont{A.}~\bibnamefont{Dollinger}},
  \bibinfo{author}{\bibfnamefont{M.}~\bibnamefont{Lochner}},
  \bibinfo{author}{\bibfnamefont{X.}~\bibnamefont{Cui}},
  \bibinfo{author}{\bibfnamefont{L.}~\bibnamefont{Patthey}},
  \bibinfo{author}{\bibfnamefont{R.}~\bibnamefont{Matzdorf}}, \bibnamefont{and}
  \bibinfo{author}{\bibfnamefont{R.}~\bibnamefont{Claessen}},
  \bibinfo{journal}{Nature Physics} \textbf{\bibinfo{volume}{7}},
  \bibinfo{pages}{776} (\bibinfo{year}{2011}).

\bibitem[{\citenamefont{Kollath et~al.}(2005)\citenamefont{Kollath,
  Schollw{\"o}ck, and Zwerger}}]{Kollath}
\bibinfo{author}{\bibfnamefont{C.}~\bibnamefont{Kollath}},
  \bibinfo{author}{\bibfnamefont{U.}~\bibnamefont{Schollw{\"o}ck}},
  \bibnamefont{and} \bibinfo{author}{\bibfnamefont{W.}~\bibnamefont{Zwerger}},
  \bibinfo{journal}{Physical review letters} \textbf{\bibinfo{volume}{95}},
  \bibinfo{pages}{176401} (\bibinfo{year}{2005}).

\bibitem[{\citenamefont{Kollath and Schollwöck}(2006)}]{Kollath2006}
\bibinfo{author}{\bibfnamefont{C.}~\bibnamefont{Kollath}} \bibnamefont{and}
  \bibinfo{author}{\bibfnamefont{U.}~\bibnamefont{Schollwöck}},
  \bibinfo{journal}{New Journal of Physics} \textbf{\bibinfo{volume}{8}},
  \bibinfo{pages}{220} (\bibinfo{year}{2006}).

\bibitem[{\citenamefont{Feiguin and Huse}(2009)}]{Feiguin2009}
\bibinfo{author}{\bibfnamefont{A.~E.} \bibnamefont{Feiguin}} \bibnamefont{and}
  \bibinfo{author}{\bibfnamefont{D.~A.} \bibnamefont{Huse}},
  \bibinfo{journal}{Phys. Rev. B} \textbf{\bibinfo{volume}{79}},
  \bibinfo{pages}{100507} (\bibinfo{year}{2009}).

\bibitem[{\citenamefont{Kim et~al.}(2006)\citenamefont{Kim, Koh, Rotenberg, Oh,
  Eisaki, Motoyama, Uchida, Tohyama, Maekawa, Shen et~al.}}]{Kim}
\bibinfo{author}{\bibfnamefont{B.}~\bibnamefont{Kim}},
  \bibinfo{author}{\bibfnamefont{H.}~\bibnamefont{Koh}},
  \bibinfo{author}{\bibfnamefont{E.}~\bibnamefont{Rotenberg}},
  \bibinfo{author}{\bibfnamefont{S.-J.} \bibnamefont{Oh}},
  \bibinfo{author}{\bibfnamefont{H.}~\bibnamefont{Eisaki}},
  \bibinfo{author}{\bibfnamefont{N.}~\bibnamefont{Motoyama}},
  \bibinfo{author}{\bibfnamefont{S.}~\bibnamefont{Uchida}},
  \bibinfo{author}{\bibfnamefont{T.}~\bibnamefont{Tohyama}},
  \bibinfo{author}{\bibfnamefont{S.}~\bibnamefont{Maekawa}},
  \bibinfo{author}{\bibfnamefont{Z.-X.} \bibnamefont{Shen}},
  \bibnamefont{et~al.}, \bibinfo{journal}{Nature Physics}
  \textbf{\bibinfo{volume}{2}}, \bibinfo{pages}{397} (\bibinfo{year}{2006}).

\bibitem[{\citenamefont{Sing et~al.}(2003)\citenamefont{Sing,
  Schwingenschl{\"o}gl, Claessen, Blaha, Carmelo, Martelo, Sacramento, Dressel,
  and Jacobsen}}]{Sing}
\bibinfo{author}{\bibfnamefont{M.}~\bibnamefont{Sing}},
  \bibinfo{author}{\bibfnamefont{U.}~\bibnamefont{Schwingenschl{\"o}gl}},
  \bibinfo{author}{\bibfnamefont{R.}~\bibnamefont{Claessen}},
  \bibinfo{author}{\bibfnamefont{P.}~\bibnamefont{Blaha}},
  \bibinfo{author}{\bibfnamefont{J.}~\bibnamefont{Carmelo}},
  \bibinfo{author}{\bibfnamefont{L.}~\bibnamefont{Martelo}},
  \bibinfo{author}{\bibfnamefont{P.}~\bibnamefont{Sacramento}},
  \bibinfo{author}{\bibfnamefont{M.}~\bibnamefont{Dressel}}, \bibnamefont{and}
  \bibinfo{author}{\bibfnamefont{C.~S.} \bibnamefont{Jacobsen}},
  \bibinfo{journal}{Physical Review B} \textbf{\bibinfo{volume}{68}},
  \bibinfo{pages}{125111} (\bibinfo{year}{2003}).

\bibitem[{\citenamefont{Lee et~al.}(2013)\citenamefont{Lee, Johnston, Moritz,
  Lee, Yi, Zhou, Schmitt, Patthey, Strocov, Kudo et~al.}}]{DeverauxPRL2013}
\bibinfo{author}{\bibfnamefont{W.~S.} \bibnamefont{Lee}},
  \bibinfo{author}{\bibfnamefont{S.}~\bibnamefont{Johnston}},
  \bibinfo{author}{\bibfnamefont{B.}~\bibnamefont{Moritz}},
  \bibinfo{author}{\bibfnamefont{J.}~\bibnamefont{Lee}},
  \bibinfo{author}{\bibfnamefont{M.}~\bibnamefont{Yi}},
  \bibinfo{author}{\bibfnamefont{K.~J.} \bibnamefont{Zhou}},
  \bibinfo{author}{\bibfnamefont{T.}~\bibnamefont{Schmitt}},
  \bibinfo{author}{\bibfnamefont{L.}~\bibnamefont{Patthey}},
  \bibinfo{author}{\bibfnamefont{V.}~\bibnamefont{Strocov}},
  \bibinfo{author}{\bibfnamefont{K.}~\bibnamefont{Kudo}}, \bibnamefont{et~al.},
  \bibinfo{journal}{Phys. Rev. Lett.} \textbf{\bibinfo{volume}{110}},
  \bibinfo{pages}{265502} (\bibinfo{year}{2013}).

\bibitem[{\citenamefont{Lee et~al.}(2014)\citenamefont{Lee, Moritz, Lee, Yi,
  Jia, Sorini, Kudo, Koike, Zhou, Monney et~al.}}]{DeverauxPRB2014}
\bibinfo{author}{\bibfnamefont{J.~J.} \bibnamefont{Lee}},
  \bibinfo{author}{\bibfnamefont{B.}~\bibnamefont{Moritz}},
  \bibinfo{author}{\bibfnamefont{W.~S.} \bibnamefont{Lee}},
  \bibinfo{author}{\bibfnamefont{M.}~\bibnamefont{Yi}},
  \bibinfo{author}{\bibfnamefont{C.~J.} \bibnamefont{Jia}},
  \bibinfo{author}{\bibfnamefont{A.~P.} \bibnamefont{Sorini}},
  \bibinfo{author}{\bibfnamefont{K.}~\bibnamefont{Kudo}},
  \bibinfo{author}{\bibfnamefont{Y.}~\bibnamefont{Koike}},
  \bibinfo{author}{\bibfnamefont{K.~J.} \bibnamefont{Zhou}},
  \bibinfo{author}{\bibfnamefont{C.}~\bibnamefont{Monney}},
  \bibnamefont{et~al.}, \bibinfo{journal}{Phys. Rev. B}
  \textbf{\bibinfo{volume}{89}}, \bibinfo{pages}{041104}
  (\bibinfo{year}{2014}).

\bibitem[{\citenamefont{Matsueda et~al.}(2006)\citenamefont{Matsueda, Tohyama,
  and Maekawa}}]{matsueda}
\bibinfo{author}{\bibfnamefont{H.}~\bibnamefont{Matsueda}},
  \bibinfo{author}{\bibfnamefont{T.}~\bibnamefont{Tohyama}}, \bibnamefont{and}
  \bibinfo{author}{\bibfnamefont{S.}~\bibnamefont{Maekawa}},
  \bibinfo{journal}{Physical Review B} \textbf{\bibinfo{volume}{74}},
  \bibinfo{pages}{241103} (\bibinfo{year}{2006}).

\bibitem[{\citenamefont{Ning et~al.}(2006)\citenamefont{Ning, Zhao, Wu, and
  Lin}}]{ning}
\bibinfo{author}{\bibfnamefont{W.-Q.} \bibnamefont{Ning}},
  \bibinfo{author}{\bibfnamefont{H.}~\bibnamefont{Zhao}},
  \bibinfo{author}{\bibfnamefont{C.-Q.} \bibnamefont{Wu}}, \bibnamefont{and}
  \bibinfo{author}{\bibfnamefont{H.-Q.} \bibnamefont{Lin}},
  \bibinfo{journal}{Physical review letters} \textbf{\bibinfo{volume}{96}},
  \bibinfo{pages}{156402} (\bibinfo{year}{2006}).

\bibitem[{\citenamefont{White and Feiguin}(2004)}]{tDMRGFeiguin}
\bibinfo{author}{\bibfnamefont{S.~R.} \bibnamefont{White}} \bibnamefont{and}
  \bibinfo{author}{\bibfnamefont{A.~E.} \bibnamefont{Feiguin}},
  \bibinfo{journal}{Phys. Rev. Lett.} \textbf{\bibinfo{volume}{93}},
  \bibinfo{pages}{076401} (\bibinfo{year}{2004}).

\bibitem[{\citenamefont{Daley et~al.}(2004)\citenamefont{Daley, Kollath,
  Schollwock, and Vidal}}]{tDMRGVidal}
\bibinfo{author}{\bibfnamefont{A.~J.} \bibnamefont{Daley}},
  \bibinfo{author}{\bibfnamefont{C.}~\bibnamefont{Kollath}},
  \bibinfo{author}{\bibfnamefont{U.}~\bibnamefont{Schollwock}},
  \bibnamefont{and} \bibinfo{author}{\bibfnamefont{G.}~\bibnamefont{Vidal}},
  \bibinfo{journal}{Journal of Statistical Mechanics: Theory and Experiment}
  \textbf{\bibinfo{volume}{2004}}, \bibinfo{pages}{P04005}
  (\bibinfo{year}{2004}).

\bibitem[{\citenamefont{Hardikar and Clay}(2007)}]{Hardikar2007}
\bibinfo{author}{\bibfnamefont{R.~P.} \bibnamefont{Hardikar}} \bibnamefont{and}
  \bibinfo{author}{\bibfnamefont{R.~T.} \bibnamefont{Clay}},
  \bibinfo{journal}{Phys. Rev. B} \textbf{\bibinfo{volume}{75}},
  \bibinfo{pages}{245103} (\bibinfo{year}{2007}).

\bibitem[{\citenamefont{Ogata and Shiba}(1990)}]{Ogata1990}
\bibinfo{author}{\bibfnamefont{M.}~\bibnamefont{Ogata}} \bibnamefont{and}
  \bibinfo{author}{\bibfnamefont{H.}~\bibnamefont{Shiba}},
  \bibinfo{journal}{Physical Review B} \textbf{\bibinfo{volume}{41}},
  \bibinfo{pages}{2326} (\bibinfo{year}{1990}).

\bibitem[{\citenamefont{Reja et~al.}(2011)\citenamefont{Reja, Yarlagadda, and
  Littlewood}}]{Littlewood2011}
\bibinfo{author}{\bibfnamefont{S.}~\bibnamefont{Reja}},
  \bibinfo{author}{\bibfnamefont{S.}~\bibnamefont{Yarlagadda}},
  \bibnamefont{and} \bibinfo{author}{\bibfnamefont{P.~B.}
  \bibnamefont{Littlewood}}, \bibinfo{journal}{Phys. Rev. B}
  \textbf{\bibinfo{volume}{84}}, \bibinfo{pages}{085127}
  (\bibinfo{year}{2011}).

\bibitem[{\citenamefont{Reja et~al.}(2012)\citenamefont{Reja, Yarlagadda, and
  Littlewood}}]{Littlewood2012}
\bibinfo{author}{\bibfnamefont{S.}~\bibnamefont{Reja}},
  \bibinfo{author}{\bibfnamefont{S.}~\bibnamefont{Yarlagadda}},
  \bibnamefont{and} \bibinfo{author}{\bibfnamefont{P.~B.}
  \bibnamefont{Littlewood}}, \bibinfo{journal}{Phys. Rev. B}
  \textbf{\bibinfo{volume}{86}}, \bibinfo{pages}{045110}
  (\bibinfo{year}{2012}).

\bibitem[{\citenamefont{Perroni et~al.}(2005)\citenamefont{Perroni, Cataudella,
  De~Filippis, and Ramaglia}}]{Perroni2005}
\bibinfo{author}{\bibfnamefont{C.~A.} \bibnamefont{Perroni}},
  \bibinfo{author}{\bibfnamefont{V.}~\bibnamefont{Cataudella}},
  \bibinfo{author}{\bibfnamefont{G.}~\bibnamefont{De~Filippis}},
  \bibnamefont{and} \bibinfo{author}{\bibfnamefont{V.~M.}
  \bibnamefont{Ramaglia}}, \bibinfo{journal}{Phys. Rev. B}
  \textbf{\bibinfo{volume}{71}}, \bibinfo{pages}{113107}
  (\bibinfo{year}{2005}).

\bibitem[{\citenamefont{Hohenadler and Assaad}(2013)}]{Assaad2013}
\bibinfo{author}{\bibfnamefont{M.}~\bibnamefont{Hohenadler}} \bibnamefont{and}
  \bibinfo{author}{\bibfnamefont{F.~F.} \bibnamefont{Assaad}},
  \bibinfo{journal}{Phys. Rev. B} \textbf{\bibinfo{volume}{87}},
  \bibinfo{pages}{075149} (\bibinfo{year}{2013}).

\bibitem[{\citenamefont{Payeur and S\'en\'echal}(2011)}]{Senechal2011}
\bibinfo{author}{\bibfnamefont{A.}~\bibnamefont{Payeur}} \bibnamefont{and}
  \bibinfo{author}{\bibfnamefont{D.}~\bibnamefont{S\'en\'echal}},
  \bibinfo{journal}{Phys. Rev. B} \textbf{\bibinfo{volume}{83}},
  \bibinfo{pages}{033104} (\bibinfo{year}{2011}).

\bibitem[{\citenamefont{Nowadnick et~al.}(2012)\citenamefont{Nowadnick,
  Johnston, Moritz, Scalettar, and Devereaux}}]{Deveraux2012}
\bibinfo{author}{\bibfnamefont{E.~A.} \bibnamefont{Nowadnick}},
  \bibinfo{author}{\bibfnamefont{S.}~\bibnamefont{Johnston}},
  \bibinfo{author}{\bibfnamefont{B.}~\bibnamefont{Moritz}},
  \bibinfo{author}{\bibfnamefont{R.~T.} \bibnamefont{Scalettar}},
  \bibnamefont{and} \bibinfo{author}{\bibfnamefont{T.~P.}
  \bibnamefont{Devereaux}}, \bibinfo{journal}{Phys. Rev. Lett.}
  \textbf{\bibinfo{volume}{109}}, \bibinfo{pages}{246404}
  (\bibinfo{year}{2012}).

\bibitem[{\citenamefont{Bauer}(2010)}]{Bauer2010}
\bibinfo{author}{\bibfnamefont{J.}~\bibnamefont{Bauer}}, \bibinfo{journal}{EPL
  (Europhysics Letters)} \textbf{\bibinfo{volume}{90}}, \bibinfo{pages}{27002}
  (\bibinfo{year}{2010}).

\bibitem[{\citenamefont{Bauer and Hewson}(2010)}]{Bauer2010b}
\bibinfo{author}{\bibfnamefont{J.}~\bibnamefont{Bauer}} \bibnamefont{and}
  \bibinfo{author}{\bibfnamefont{A.~C.} \bibnamefont{Hewson}},
  \bibinfo{journal}{Phys. Rev. B} \textbf{\bibinfo{volume}{81}},
  \bibinfo{pages}{235113} (\bibinfo{year}{2010}).

\bibitem[{\citenamefont{Kumar and van~den Brink}(2008)}]{vandenBrink2008}
\bibinfo{author}{\bibfnamefont{S.}~\bibnamefont{Kumar}} \bibnamefont{and}
  \bibinfo{author}{\bibfnamefont{J.}~\bibnamefont{van~den Brink}},
  \bibinfo{journal}{Phys. Rev. B} \textbf{\bibinfo{volume}{78}},
  \bibinfo{pages}{155123} (\bibinfo{year}{2008}).

\bibitem[{\citenamefont{Barone et~al.}(2008)\citenamefont{Barone, Raimondi,
  Capone, Castellani, and Fabrizio}}]{Fabrizio2008}
\bibinfo{author}{\bibfnamefont{P.}~\bibnamefont{Barone}},
  \bibinfo{author}{\bibfnamefont{R.}~\bibnamefont{Raimondi}},
  \bibinfo{author}{\bibfnamefont{M.}~\bibnamefont{Capone}},
  \bibinfo{author}{\bibfnamefont{C.}~\bibnamefont{Castellani}},
  \bibnamefont{and} \bibinfo{author}{\bibfnamefont{M.}~\bibnamefont{Fabrizio}},
  \bibinfo{journal}{Phys. Rev. B} \textbf{\bibinfo{volume}{77}},
  \bibinfo{pages}{235115} (\bibinfo{year}{2008}).

\bibitem[{\citenamefont{Fehske et~al.}(2008)\citenamefont{Fehske, Hager, and
  Jeckelmann}}]{Fehske2008}
\bibinfo{author}{\bibfnamefont{H.}~\bibnamefont{Fehske}},
  \bibinfo{author}{\bibfnamefont{G.}~\bibnamefont{Hager}}, \bibnamefont{and}
  \bibinfo{author}{\bibfnamefont{E.}~\bibnamefont{Jeckelmann}},
  \bibinfo{journal}{EPL (Europhysics Letters)} \textbf{\bibinfo{volume}{84}},
  \bibinfo{pages}{57001} (\bibinfo{year}{2008}).

\bibitem[{\citenamefont{Tezuka et~al.}(2007)\citenamefont{Tezuka, Arita, and
  Aoki}}]{Tezuka2007}
\bibinfo{author}{\bibfnamefont{M.}~\bibnamefont{Tezuka}},
  \bibinfo{author}{\bibfnamefont{R.}~\bibnamefont{Arita}}, \bibnamefont{and}
  \bibinfo{author}{\bibfnamefont{H.}~\bibnamefont{Aoki}},
  \bibinfo{journal}{Phys. Rev. B} \textbf{\bibinfo{volume}{76}},
  \bibinfo{pages}{155114} (\bibinfo{year}{2007}).

\bibitem[{\citenamefont{Ejima and Fehske}(2010)}]{Fehske2010}
\bibinfo{author}{\bibfnamefont{S.}~\bibnamefont{Ejima}} \bibnamefont{and}
  \bibinfo{author}{\bibfnamefont{H.}~\bibnamefont{Fehske}},
  \bibinfo{journal}{Journal of Physics: Conference Series}
  \textbf{\bibinfo{volume}{200}}, \bibinfo{pages}{012031}
  (\bibinfo{year}{2010}).

\bibitem[{\citenamefont{Ejima and Fehske}(2009)}]{Fehske2009}
\bibinfo{author}{\bibfnamefont{S.}~\bibnamefont{Ejima}} \bibnamefont{and}
  \bibinfo{author}{\bibfnamefont{H.}~\bibnamefont{Fehske}},
  \bibinfo{journal}{EPL (Europhysics Letters)} \textbf{\bibinfo{volume}{87}},
  \bibinfo{pages}{27001} (\bibinfo{year}{2009}).

\bibitem[{\citenamefont{Jeckelmann and White}(1998)}]{WhiteJeckelmann98}
\bibinfo{author}{\bibfnamefont{E.}~\bibnamefont{Jeckelmann}} \bibnamefont{and}
  \bibinfo{author}{\bibfnamefont{S.~R.} \bibnamefont{White}},
  \bibinfo{journal}{Phys. Rev. B} \textbf{\bibinfo{volume}{57}},
  \bibinfo{pages}{6376} (\bibinfo{year}{1998}).

\bibitem[{\citenamefont{Zhang et~al.}(1999)\citenamefont{Zhang, Jeckelmann, and
  White}}]{WhiteJeckelmann99}
\bibinfo{author}{\bibfnamefont{C.}~\bibnamefont{Zhang}},
  \bibinfo{author}{\bibfnamefont{E.}~\bibnamefont{Jeckelmann}},
  \bibnamefont{and} \bibinfo{author}{\bibfnamefont{S.~R.} \bibnamefont{White}},
  \bibinfo{journal}{Phys. Rev. B} \textbf{\bibinfo{volume}{60}},
  \bibinfo{pages}{14092} (\bibinfo{year}{1999}).

\bibitem[{\citenamefont{Cataudella et~al.}(2004)\citenamefont{Cataudella,
  De~Filippis, Martone, and Perroni}}]{Perroni2004}
\bibinfo{author}{\bibfnamefont{V.}~\bibnamefont{Cataudella}},
  \bibinfo{author}{\bibfnamefont{G.}~\bibnamefont{De~Filippis}},
  \bibinfo{author}{\bibfnamefont{F.}~\bibnamefont{Martone}}, \bibnamefont{and}
  \bibinfo{author}{\bibfnamefont{C.~A.} \bibnamefont{Perroni}},
  \bibinfo{journal}{Phys. Rev. B} \textbf{\bibinfo{volume}{70}},
  \bibinfo{pages}{193105} (\bibinfo{year}{2004}).

\bibitem[{\citenamefont{Manmana et~al.}(2005)\citenamefont{Manmana, Muramatsu,
  and Noack}}]{tDMRGManmana}
\bibinfo{author}{\bibfnamefont{S.~R.} \bibnamefont{Manmana}},
  \bibinfo{author}{\bibfnamefont{A.}~\bibnamefont{Muramatsu}},
  \bibnamefont{and} \bibinfo{author}{\bibfnamefont{R.~M.} \bibnamefont{Noack}},
  \bibinfo{journal}{AIP Conference Proceedings} \textbf{\bibinfo{volume}{789}}
  (\bibinfo{year}{2005}).

\bibitem[{\citenamefont{Schmitteckert}(2004)}]{tDMRGSchmitt}
\bibinfo{author}{\bibfnamefont{P.}~\bibnamefont{Schmitteckert}},
  \bibinfo{journal}{Phys. Rev. B} \textbf{\bibinfo{volume}{70}},
  \bibinfo{pages}{121302} (\bibinfo{year}{2004}).

\bibitem[{\citenamefont{Cazalilla and Marston}(2002)}]{cazalilla2002}
\bibinfo{author}{\bibfnamefont{M.}~\bibnamefont{Cazalilla}} \bibnamefont{and}
  \bibinfo{author}{\bibfnamefont{J.}~\bibnamefont{Marston}},
  \bibinfo{journal}{Physical review letters} \textbf{\bibinfo{volume}{88}},
  \bibinfo{pages}{256403} (\bibinfo{year}{2002}).

\bibitem[{\citenamefont{Cazalilla and Marston}(2003)}]{cazalilla2003}
\bibinfo{author}{\bibfnamefont{M.}~\bibnamefont{Cazalilla}} \bibnamefont{and}
  \bibinfo{author}{\bibfnamefont{J.}~\bibnamefont{Marston}},
  \bibinfo{journal}{Physical Review Letters} \textbf{\bibinfo{volume}{91}},
  \bibinfo{pages}{049702} (\bibinfo{year}{2003}).

\bibitem[{\citenamefont{Luo et~al.}(2003)\citenamefont{Luo, Xiang, and
  Wang}}]{luo}
\bibinfo{author}{\bibfnamefont{H.}~\bibnamefont{Luo}},
  \bibinfo{author}{\bibfnamefont{T.}~\bibnamefont{Xiang}}, \bibnamefont{and}
  \bibinfo{author}{\bibfnamefont{X.}~\bibnamefont{Wang}},
  \bibinfo{journal}{Physical review letters} \textbf{\bibinfo{volume}{91}},
  \bibinfo{pages}{49701} (\bibinfo{year}{2003}).

\bibitem[{\citenamefont{Zheng et~al.}(1989)\citenamefont{Zheng, Feinberg, and
  Avignon}}]{Zheng}
\bibinfo{author}{\bibfnamefont{H.}~\bibnamefont{Zheng}},
  \bibinfo{author}{\bibfnamefont{D.}~\bibnamefont{Feinberg}}, \bibnamefont{and}
  \bibinfo{author}{\bibfnamefont{M.}~\bibnamefont{Avignon}},
  \bibinfo{journal}{Phys. Rev. B} \textbf{\bibinfo{volume}{39}},
  \bibinfo{pages}{9405} (\bibinfo{year}{1989}).

\bibitem[{\citenamefont{Perroni et~al.}(2003)\citenamefont{Perroni, Cataudella,
  De~Filippis, Iadonisi, Marigliano~Ramaglia, and Ventriglia}}]{Perroni2003}
\bibinfo{author}{\bibfnamefont{C.~A.} \bibnamefont{Perroni}},
  \bibinfo{author}{\bibfnamefont{V.}~\bibnamefont{Cataudella}},
  \bibinfo{author}{\bibfnamefont{G.}~\bibnamefont{De~Filippis}},
  \bibinfo{author}{\bibfnamefont{G.}~\bibnamefont{Iadonisi}},
  \bibinfo{author}{\bibfnamefont{V.}~\bibnamefont{Marigliano~Ramaglia}},
  \bibnamefont{and}
  \bibinfo{author}{\bibfnamefont{F.}~\bibnamefont{Ventriglia}},
  \bibinfo{journal}{Phys. Rev. B} \textbf{\bibinfo{volume}{67}},
  \bibinfo{pages}{214301} (\bibinfo{year}{2003}).

\bibitem[{\citenamefont{Lang and Firsov}(1963)}]{LangFirsov}
\bibinfo{author}{\bibfnamefont{I.~J.} \bibnamefont{Lang}} \bibnamefont{and}
  \bibinfo{author}{\bibfnamefont{Y.~A.} \bibnamefont{Firsov}},
  \bibinfo{journal}{Sov. Phys. JETP} \textbf{\bibinfo{volume}{16}},
  \bibinfo{pages}{1301} (\bibinfo{year}{1963}).

\bibitem[{\citenamefont{Penc et~al.}(1997)\citenamefont{Penc, Hallberg, Mila,
  and Shiba}}]{Penc1997b}
\bibinfo{author}{\bibfnamefont{K.}~\bibnamefont{Penc}},
  \bibinfo{author}{\bibfnamefont{K.}~\bibnamefont{Hallberg}},
  \bibinfo{author}{\bibfnamefont{F.}~\bibnamefont{Mila}}, \bibnamefont{and}
  \bibinfo{author}{\bibfnamefont{H.}~\bibnamefont{Shiba}},
  \bibinfo{journal}{Phys. Rev. B} \textbf{\bibinfo{volume}{55}},
  \bibinfo{pages}{15475} (\bibinfo{year}{1997}).

\end{thebibliography}

\end{document}